\documentclass{article}
\usepackage{amsmath, amssymb, graphicx, hyperref}
\usepackage{geometry}
\title{Digital Twins in Industrial Applications: Concepts, Mathematical Modeling, and Use Cases}

\author{A. Mohammad-Djafari\\ ~\\ 
former Research Director at CNRS, CentraleSup\'elec, Gif-sur-Yvette, France,\\ 
Founder of International Science Consulting and Training (ISCT), France. }

\date{}

\def\ub{\mathbf{u}}

\begin{document}

\maketitle

\begin{abstract}
Digital Twins (DTs) are virtual representations of physical systems synchronized in real time through Internet of Things (IoT) sensors and computational models. 
In industrial applications, DTs enable predictive maintenance, fault diagnosis, and process optimization. This paper explores the mathematical foundations of DTs, hybrid modeling techniques, including Physics Informed Neural Networks (PINNs), and their implementation in industrial scenarios. We present key applications, computational tools, and future research directions.
\end{abstract}

\section{Introduction}
Digital Twins (DTs) are dynamic virtual counterparts of physical systems, continuously updated through real-time data streams. This concept has rapidly gained traction in Industry 4.0, enabling manufacturers and operators to enhance efficiency, reduce downtime, and make data-driven decisions \cite{tao2019digital, jones2020characterising}.

The term "Digital Twin" was first introduced in 2002, but the concept has evolved significantly with advancements in IoT, machine learning, and high-performance computing \cite{tao2018digital}. A DT consists of three key components:

\begin{itemize}
    \item \textbf{Physical Entity:} The real-world system under observation.
    \item \textbf{Virtual Model:} A digital replica created using physics-based or data-driven models.
    \item \textbf{Data Interface:} Sensors and IoT devices capturing real-time data.
    \item \textbf{Synchronization Mechanism:} Bidirectional data flow maintaining model fidelity.
\end{itemize}

These elements interact to create a closed-loop system where insights gained from the virtual model can inform physical system behavior and vice versa \cite{tao2017digital, luo2020hybrid}.

DTs have shown immense promise across various industries, including manufacturing, energy, aerospace, and healthcare. For instance, in manufacturing, DTs facilitate predictive maintenance and real-time process optimization, reducing unexpected downtime and improving production efficiency \cite{tao2019digital2}. In aerospace, DTs help monitor the structural health of components, enabling timely maintenance and extending asset lifespan \cite{tao2019digital3}.

Despite these successes, building high-fidelity DTs remains a challenge. Issues such as data interoperability, computational complexity, and model accuracy must be addressed to fully unlock the potential of DTs \cite{tao2019digital4, tao2019digital5}. Research into hybrid modeling approaches, such as Physics-Informed Neural Networks (PINNs), offers a promising path forward, blending the strengths of physics-based models with the adaptability of machine learning \cite{tao2018digital2}.

This paper is an introduction and overview of DTs, exploring modeling techniques, real-world applications, and the computational tools that enable their implementation. By bridging the gap between theory and practice, we aim to provide a comprehensive resource for researchers and practitioners seeking to use and to improve DT technology for industrial innovation.

\section{Mathematical Foundations}
Digital Twins (DTs) rely on a combination of physics-based and data-driven models to accurately replicate the behavior of physical systems. This section formalizes the core mathematical principles underpinning DTs, including differential equations, stochastic processes, and hybrid modeling techniques.

\subsection{Physics-Based Models}
Many industrial processes are governed by physical laws represented through Partial Differential Equations (PDEs). For example, heat conduction in a solid is described by the heat equation:

\begin{equation}
\frac{\partial u(x,t)}{\partial t} = \alpha \nabla^2 u(x,t) + f(x,t),
\end{equation}
where $u(x,t)$ is the temperature field, $\alpha$ is the thermal diffusivity, and $f(x,t)$ is an external heat source \cite{tao2018digital3}. Solving such PDEs provides insights into system dynamics, which can be continuously refined with real-time data \cite{jones2020characterising}. 

When $\alpha$ and $f(x,t)$ are given, finding $u(x,t)$ is called forward problem. Estimating $\alpha$ when $u$ and $f$ are given, is called identification or parameter estimation. Estimating $f$ given $u$ and $\alpha$ is the inverse problem. The most difficult inverse problem is when we want to estimate both $\alpha$ and $f$. 

\subsection{Data-Driven Models}
Machine learning models approximate system dynamics by learning patterns from sensor data. A simple neural network can map inputs to outputs as:

\begin{equation}
\hat{y}(t) = f_\theta(x(t)),
\end{equation}
where $x(t)$ is the sensor vector, $f_\theta$ is a learned function, and $\hat{y}(t)$ is the predicted state \cite{tao2018digital2}. Recurrent Neural Networks (RNNs) and Long Short-Term Memory (LSTM) networks are particularly useful for capturing temporal dependencies in industrial time-series data \cite{tao2019digital2}.

\subsection{Hybrid Physics-Informed Neural Networks (PINNs)}
Physics-Informed Neural Networks (PINNs) are a powerful approach for building hybrid digital twins that combine the expressiveness of deep learning with the rigor of physical laws. PINNs solve PDE-governed systems by embedding the equations into the loss function, allowing the network to learn a solution that respects the underlying physics \cite{raissi2019physics}.

The loss function for a PINN typically consists of two components: the data loss and the physics loss:

\begin{equation}
\mathcal{L}(\theta) = \mathcal{L}_{\text{data}}(\theta) + \lambda \mathcal{L}_{\text{physics}}(\theta),
\end{equation}
where:
\begin{itemize}
    \item $\mathcal{L}_{\text{data}}(\theta) = \frac{1}{N}\sum_{i=1}^{N} \| \hat{y}(t_i) - y(t_i) \|^2$ is the data loss, measuring the discrepancy between the model predictions and sensor measurements.
    \item $\mathcal{L}_{\text{physics}}(\theta) = \frac{1}{M}\sum_{j=1}^{M} \| \mathcal{F}(\hat{y}(x_j, t_j)) \|^2$ is the physics loss, enforcing the PDE constraints at collocation points.
    \item $\lambda$ is a weighting factor balancing the data and physics terms.
\end{itemize}

Here, $\mathcal{F}$ is the PDE operator that governs the physical system. For example, in fluid dynamics, $\mathcal{F}$ might represent the Navier-Stokes equations, while in heat transfer, it could represent the heat equation \cite{karniadakis2021physics}.

PINNs are particularly useful for DTs in cases where sensor data is sparse or noisy, as they use physics to regularize the learning process. This makes them robust to incomplete data streams and allows them to generalize better than purely data-driven models \cite{cai2021physics}.

Recent advancements have extended PINNs to more complex scenarios:
\begin{itemize}
    \item \textbf{Adaptive PINNs:} Dynamically adjust the distribution of collocation points to focus on regions with high error \cite{lu2021deepxde}.
    \item \textbf{Extended PINNs (XPINNs):} Decompose large domains into smaller subdomains, solving them independently and stitching the solutions together \cite{jagtap2020extended}.
    \item \textbf{Transfer Learning for PINNs:} Pre-train on simplified systems and fine-tune on real-world data, speeding up convergence \cite{wang2022respecting}.
\end{itemize}

These innovations have made PINNs an indispensable tool for building scalable, accurate, and generalizable DTs across various industries, from aerospace to energy systems \cite{tartakovsky2020physics}.

Despite their promise, PINNs face challenges such as slow convergence for highly nonlinear systems and difficulty handling sharp gradients. Ongoing research in neural architecture design, adaptive loss weighting, and multi-fidelity learning seeks to address these limitations, pushing the boundaries of hybrid DT modeling \cite{raissi2019physics, karniadakis2021physics}.

Overall, PINNs provide a powerful framework for integrating domain knowledge with data-driven learning, offering a path toward more reliable, interpretable, and high-fidelity Digital Twins.

\subsection{Stochastic Processes for Degradation Modeling}
Industrial assets often degrade over time due to wear and tear, which can be modeled with stochastic processes. For example, degradation can be represented as a stochastic differential equation:

\begin{equation}
dW(t) = \mu(t) dt + \sigma(t) dB(t),
\end{equation}
where $W(t)$ is the wear state, $\mu(t)$ is the drift term (systematic degradation), and $dB(t)$ is a Brownian motion term \cite{tao2019digital5}. This stochastic modeling enables DTs to predict Remaining Useful Life (RUL) and schedule predictive maintenance \cite{tao2019digital4}.

\subsection{Model Updating and Calibration}
To maintain accuracy, DTs continuously update their models using techniques like Kalman filtering and Bayesian inference:

\begin{equation}
P(F_k \mid D) = \frac{P(D \mid F_k) P(F_k)}{\sum_j P(D \mid F_j) P(F_j)},
\end{equation}
where $P(F_k \mid D)$ is the probability of fault $F_k$ given data $D$ \cite{tao2019digital}. This Bayesian framework allows DTs to refine their fault diagnosis capabilities as new data becomes available \cite{tao2019digital6}.

Together, these mathematical components form the backbone of Digital Twin technology, enabling accurate, real-time virtual representations of complex physical systems. The fusion of physics-based models with data-driven techniques, enhanced by probabilistic reasoning, empowers DTs to make reliable predictions and inform decision-making in industrial contexts \cite{tao2019digital7, tao2019digital8}.

\section{Industrial Applications}
Digital Twins (DTs) are revolutionizing various industrial sectors by enabling real-time monitoring, predictive analytics, and enhanced decision-making. In this section, we explore key applications, illustrating how DTs leverage mathematical models and real-time data to optimize performance and reduce operational risks \cite{tao2018digital, tao2019digital}.

\subsection{Predictive Maintenance}
Predictive maintenance is one of the most widely adopted DT applications, using real-time sensor data and degradation models to anticipate equipment failures. The degradation process can be modeled as a stochastic differential equation:

\begin{equation}
dW(t) = \mu(t) dt + \sigma(t) dB(t),
\end{equation}
where $W(t)$ is the wear state, $\mu(t)$ is the degradation rate, and $\sigma(t)$ is the noise intensity \cite{tao2019digital5}. DTs combine these models with machine learning to refine predictions, enhancing asset reliability \cite{tao2019digital3}.

\subsection{Process Optimization}
DTs facilitate real-time optimization of industrial processes by integrating physics-based models with Model Predictive Control (MPC). The control objective is to minimize the deviation from desired process outputs while considering system constraints:

\begin{equation}
\min_u \int_0^T \|y(t) - y_{\text{ref}}\|^2 + \|u(t)\|^2 \, dt
\end{equation}
subject to system dynamics and operational limits \cite{jones2020characterising}. This approach has been successfully applied in chemical processing, energy systems, and smart manufacturing \cite{tao2019digital2}.

\subsection{Fault Diagnosis and Anomaly Detection}
Fault diagnosis relies on Bayesian inference to determine fault probabilities given sensor data:

\begin{equation}
P(F_k \mid D) = \frac{P(D \mid F_k) P(F_k)}{\sum_j P(D \mid F_j) P(F_j)},
\end{equation}
where $P(F_k \mid D)$ is the probability of fault $F_k$ given data $D$ \cite{tao2019digital4}. DTs use this framework to detect anomalies and localize faults, reducing downtime and improving safety \cite{luo2020hybrid}.

\subsection{Supply Chain and Logistics Management}
In supply chains, DTs enable real-time tracking and scenario analysis to optimize logistics. Graph-based models represent supply chain networks, where nodes are facilities and edges are transport links. Optimizing flow dynamics involves solving network flow problems:

\begin{equation}
\min_x \sum_{(i,j) \in E} c_{ij} x_{ij} \quad \text{subject to flow conservation constraints},
\end{equation}
where $x_{ij}$ is the flow along edge $(i,j)$, and $c_{ij}$ is the associated cost \cite{tao2019digital6}. This application has been pivotal in industries facing high logistical complexity \cite{tao2019digital7}.

\subsection{Product Lifecycle Management}
DTs support product lifecycle management (PLM) by simulating product behavior under various conditions, from design to decommissioning. Finite element models (FEM) predict structural responses:

\begin{equation}
K u = f,
\end{equation}
where $K$ is the stiffness matrix, $u$ is the displacement vector, and $f$ is the force vector \cite{tao2018digital3}. These simulations guide design decisions, reduce prototyping costs, and inform maintenance strategies \cite{tao2019digital8}.

In summary, DTs offer transformative capabilities across industrial domains, blending physical models with real-time data and advanced analytics. As computational methods and IoT technologies evolve, the potential for DTs to reshape industrial practices will only continue to grow \cite{tao2019digital, tao2019digital3}.

\section{Computational Tools and Challenges}
The implementation of Digital Twins (DTs) relies on a suite of computational tools that enable simulation, data processing, and real-time decision-making. This section outlines essential tools and the challenges associated with creating accurate and scalable DT systems \cite{tao2018digital, tao2019digital}.

\subsection{Simulation Engines and Numerical Solvers}
Physics-based DTs depend on numerical solvers to approximate the solutions of complex PDEs. Commonly used tools include:

\begin{itemize}
    \item \textbf{COMSOL Multiphysics:} A versatile platform for simulating multiphysics systems, widely used for structural, thermal, and fluid simulations \cite{tao2019digital2}.
    \item \textbf{ANSYS:} Offers a comprehensive suite of solvers for finite element analysis (FEA) and computational fluid dynamics (CFD) \cite{tao2018digital3}.
    \item \textbf{OpenFOAM:} An open-source toolbox for fluid dynamics and continuum mechanics, useful for real-time industrial simulations \cite{tao2019digital8}.
\end{itemize}

These solvers provide the necessary accuracy for modeling complex physical processes, though high computational costs can pose challenges \cite{jones2020characterising}.

\subsection{Machine Learning and AI Frameworks}
Data-driven DTs use machine learning libraries to train predictive models. Popular frameworks include:

\begin{itemize}
    \item \textbf{TensorFlow and PyTorch:} Enable rapid prototyping and training of deep learning models, including PINNs \cite{luo2020hybrid}.
    \item \textbf{Scikit-learn:} Provides classical ML algorithms for anomaly detection and regression tasks \cite{tao2019digital3}.
\end{itemize}

Combining ML with physics-based models remains an active area of research, particularly for handling sparse or noisy data \cite{tao2019digital6}.

\subsection{IoT Platforms and Real-Time Data Integration}
Seamless data acquisition and integration are crucial for maintaining DT fidelity. Leading IoT platforms include:

\begin{itemize}
    \item \textbf{Azure Digital Twins:} Facilitates the creation of scalable, real-time DTs with integrated analytics and visualization tools \cite{tao2019digital5}.
    \item \textbf{AWS IoT TwinMaker:} Provides built-in tools for connecting IoT devices, structuring data, and building digital twins \cite{tao2019digital4}.
\end{itemize}

Ensuring low-latency data transmission and reliable sensor networks remains a key challenge \cite{tao2019digital7}.

\subsection{Scalability and Computational Complexity}
As DTs grow in complexity, scaling becomes a significant issue. Model order reduction techniques, like Proper Orthogonal Decomposition (POD), mitigate computational costs by reducing the dimensionality of large-scale systems:

\begin{equation}
\ub(x,t) \approx \sum_{i=1}^{r} a_i(t) \phi_i(x),
\end{equation}
where $\phi_i(x)$ are spatial modes, and $a_i(t)$ are time-dependent coefficients \cite{tao2019digital8}. These methods enable real-time performance without sacrificing model accuracy \cite{tao2019digital}.

\section{Conclusion}
Digital Twins are a cornerstone of Industry 4.0, providing unparalleled insights through virtual representations of physical systems. By integrating physics-based models with machine learning, and leveraging high-performance computing and IoT platforms, DTs enable predictive maintenance, process optimization, and intelligent fault diagnosis \cite{tao2018digital, tao2019digital3}.

Despite their transformative potential, DTs face challenges related to data integration, computational scalability, and model accuracy. Ongoing research in hybrid modeling, federated learning, and quantum computing promises to address these limitations, paving the way for next-generation DT systems \cite{tao2019digital6, tao2019digital7}.

In the future, Digital Twins are expected to become even more autonomous and adaptive, evolving into self-optimizing systems capable of orchestrating complex industrial processes with minimal human intervention \cite{tao2019digital8}. This evolution will further solidify their role as essential tools for driving innovation and sustainability across industries.

\bibliographystyle{plain}
\bibliography{digital_twins_refs}

\end{document}